\documentclass{amsart}          

\usepackage{amsmath,amsthm}     
\usepackage{amssymb}            

\newtheorem{thm}{Theorem}[section]
\newtheorem{theorem}[thm]{Theorem}

\theoremstyle{definition}  

\DeclareMathOperator{\Orb}{Orb\, }

\DeclareMathOperator{\supp}{supp\, } 
\DeclareMathOperator{\tr}{tr} 
  
\DeclareMathOperator{\spa}{span\, }

\DeclareMathOperator{\rank}{rank\,}

\DeclareMathOperator{\id}{id\,}
\DeclareMathOperator{\dom}{domain\,}
\DeclareMathOperator{\HS}{HS\,} \DeclareMathOperator{\Ad}{Ad\,}
\DeclareMathOperator{\ad}{ad\,}
\DeclareMathOperator{\goi}{\left(\int_\rightarrow^G\right)_0^1}
\newcommand{\g}{\mathfrak{g}}

\newcommand{\C}{\mathbf{C}}
\newcommand{\R}{\mathbf{R}}

\newcommand{\tensor}{\otimes}

\newcommand{\bigunion}{\bigcup}

\newcommand{\url}[1]{\texttt{#1}}

\begin{document}

\title[Noncommutative quantum gauge theories]{Functional integral approach to quantum gauge field theories on a noncommutative
space-time}

\author[Nahum Zobin]{Nahum Zobin}

\email{zobin@math.wm.edu}

\address{Department of Mathematics\\College of William and Mary \\ Williamsburg
\\VA 23187-8795}






\maketitle


\section{Introduction}
We discuss a functional integral approach to construction of
Lorentz-covariant quantum gauge theories on a noncommutative
space-time. There have been quite a number of work in this
direction, mostly using various Moyal-type star products to
construct Lagrangians. One of the most influential works was that
of Seiberg and Witten \cite{SW}, where they, among many other
things, noted that simple problems of evolution of a string in a
background force field invariably leads to some kind of
noncommutativity of space-time coordinates. The type of
noncommutativity they were using led to violation of the Lorentz
covariance. There was a lot of works discussing these violations
and attempting to fix this problem. In our paper \cite{CCZ} we
proposed a version of the Moyal-type approach based on a group
earlier used by Doplicher-Fredenhagen-Roberts \cite{DFR} for other
reasons. We came to the this group by contracting the group
$SO(4,1)$ used by Snyder \cite{S} to treat noncommutative
space-time.

Though our approach allowed to avoid Lorentz-covariance
violations, there were other problems that were dealt with in
quite artificial ways -- the most important being related to
treating gauge fields in the noncommutative setting.

Reflecting on this circle of ideas we were led to consideration of
a functional integral methods, based on deformation of a
commutative group to a family of noncommutative ones. This
approach leads to a natural way of constructing functions of
fields on a noncommutative space-time, and the formulas suggest
that the space of probability measures on the classical space-time
is a natural (though infinite-dimensional) noncommutative analogue
of the classical space-time. The idea is to view measures as "fat
points" which are indistinguishable from usual points of the
space-time if the scale is not small enough.

We would like to note that this idea seems to be very close to the
ideology of string theory. To some extent, it follows from our
considerations that non-commutativity of space-time invariably
leads to a version of string field theory.

This article is a preliminary exposition of our results, we plan
to write a more comprehensive paper, where we shall explore the
connections with other approaches.

I would like to thank Carl Carlson, Chris Carone, Josh Ehrlich and
Gene Tracy for valuable and illuminating discussions of many
questions arising in relation to the problems studied in this
article.

\section{Doplicher-Fredenhagen-Roberts algebra}

Consider the following $10$-dimensional Lie algebra $\g_\epsilon =
\R^4 \oplus (\R^4\wedge \R^4),\, \, \epsilon > 0,$ with the
bracket
$$\forall\, x,y \in \R^4,\, \, \, a,b \in \R^4\wedge \R^4\quad [(x,a),(y,b)] = (0, \epsilon x\wedge
y).$$ It is easy to check that this bracket indeed defines a Lie
algebra structure on $\g_\epsilon,$ and that
$\mathfrak{z}_{\g_\epsilon} = \{ (0,a): a\in \R^4\wedge \R^4\}$ is
the center of this algebra. This Lie algebra is a two step
nilpotent Lie algebra.

The linear space $\R^4 \oplus (\R^4\wedge \R^4)$ has another Lie
algebra structure -- that of a commutative Lie algebra, with all
brackets equal to zero. We denote this commutative Lie algebra
$\g_0.$

Let $G_\epsilon,\, \, \epsilon > 0,$ (resp., $G_0$) denote the
connected simply connected Lie group having $\g_\epsilon$ (resp.,
$\g_0$) as its Lie algebra. Obviously, $G_\epsilon$ and $G_0$
coincide with $\R^{10}$ as sets, and the group operations are
given by the following formulas:

in the group $G_\epsilon$

$$\forall\, X,Y \in \R^4,\, \, A,B \in \R^4\wedge \R^4\quad
(X,A)\diamondsuit (Y,B) = (X + Y, A + B +  \frac {\epsilon}2
X\wedge Y),$$

in the group $G_0$

$$\forall\, X,Y \in \R^4,\, \, A,B \in \R^4\wedge \R^4\quad
(X,A)+(Y,B) = (X + Y, A + B).$$

It is easy to see that the following is true:

(i) both groups have the same neutral element $(0,0),$

(ii) both groups have the same group inversion operation $(X,A)
\mapsto (-X, -A),$

(iii) both groups have the same left- and right-invariant Haar
measure -- the Lebesgue measure $dH(X,A) = d^4Xd^6A.$

The group $G_0$ is obviously commutative, while the group
$G_\epsilon,\, \, \epsilon > 0,$ is not, though $G_\epsilon$ is
very close to a commutative group -- it is a unipotent group. In
particular, its finite dimensional irreducible representations
have to be one-dimensional. The group $G_\epsilon$ has a huge
center
$$Z_{G_\epsilon} = \{ (0,A): A\in \R^4\wedge \R^4 \}.$$

One can compute the sets of (equivalence classes) of unitary
irreducible representations of both groups.

Obviously,
$$\widehat{G_0} = \R^{*4} \oplus SS(\R^4),$$ where $SS(\R^4)$
denotes the set of skew symmetric bilinear forms on $\R^4,$ the
dual space to $\R^4 \wedge \R^4.$ Since $G_0$ is commutative, all
irreducible unitary representations are one-dimensional, and they
are all given by the characters $$\chi_{(\phi, \Phi)}(X,A) = \exp
\, \, i(\phi(X) + \Phi(A)),\, \, (X,A) \in G_0,\, \, \phi \in
\R^{*4},\, \, \, \Phi \in SS(\R^4).$$ The Plancherel measure
$dP_0(\phi,\Phi)$ on $\widehat{G_0}$ is an appropriately  scaled
Lebesgue measure on $\R^{*4} \oplus SS(\R^4),$ more precisely,
$$dP_0(\phi,\Phi) = (2\pi)^{-10} d^4\phi d^6 \Phi.$$

A description of unitary irreducible representations of
$G_\epsilon$ is less obvious, but it can be directly derived,
e.g., using the Kirillov's Orbit Method, see, e.g., \cite {K, ZS,
BR}). We actually do not need this description, therefore we only
very briefly present it here.

One can show that
$$\widehat{G_\epsilon} = \R^{*4} \bigunion SS(\R^4).$$ The $\R^{*4}$ part
describes one dimensional representations of $G$, given by the
formula
$$\chi_\phi(X,A) = \exp i \phi(X),\quad \phi \in \R^{*4},$$ while
the $SS(\R^4)$ part describes the infinite-dimensional
representations. The Plancherel measure $dP_\epsilon$ is supported
on the $SS(\R^4)$ part, and it can be shown that
$$dP_\epsilon (\Phi) = \sqrt{\det \Phi}\, \,  d^6\Phi.$$ This means that the Plancherel
measure is supported by the nondegenerate skew symmetric bilinear
forms, i.e., by symplectic forms on $\R^4.$ Since we are going to
consider the Fourier transform on $G$, we are really interested
only in the unitary irreducible representations associated with
symplectic forms. To get these representations, consider a
symplectic form $\Phi$, then choose two complementary
$2$-dimensional $\Phi$-Lagrangian subspaces $l,l'$ in $\R^4,$ and
consider the usual (infinite-dimensional) irreducible Heisenberg
representation in $L_2(l),$ see, e.g., \cite{T}.

\section{Fourier transform and Quantization mapping}

Consider a function $f \in L_2(G_0, dH) = L_2(\R^{10},d^{10}x).$
Let $$\mathfrak{F}_0: L_2(G_0, dH) \rightarrow L_2(\widehat{G_0},
dP_0)$$ be the Fourier transform on the commutative group $G_0,$
which is simply the usual Fourier transform on $\R^{10}:$
$$(\mathfrak{F}_0 f)(\phi, \Phi) = \int_{G_0}
\chi_{(\phi,\Phi)}(X,A) f(X,A) dH(X,A)$$ $$ = \int_{\R^{10}} \exp
(i(\phi(X) + \Phi(A))) f(X,A) d^4Xd^6A.$$ This mapping is a
unitary operator, and its inverse is well known.

We can also consider the mapping $$\mathfrak{F}: L_2(G_\epsilon,
dH) \rightarrow L_2(\widehat{G_\epsilon}, \Lambda(G_\epsilon);
dP_\epsilon)$$ -- the Fourier transform on the noncommutative
group $G_\epsilon$, given by the formula
$$(\mathfrak{F} f)(\rho) = \int_{G_\epsilon} \rho ((X,A)) f(X,A) dH(X,A) =
\int_{\R^{10}}  \rho ((X,A)) f(X,A) d^4X d^6A.$$ This mapping
sends a function $f$ to a section $F$ of the {\bf dual bundle}
$$\Lambda(G_\epsilon) = \bigunion_{\rho \in \widehat{G_\epsilon}} HS (H_\rho),$$
where $H_\rho$ is the Hilbert space of the unitary irreducible
representation $\rho \in \widehat{G_\epsilon}$, and $HS(H_\rho)$
is the space of Hilbert-Schmidt operators in this space. So, $F =
\mathfrak{F} f$ is a function on $\widehat{G_\epsilon},$ whose
value at $\rho \in \widehat{G_\epsilon}$ is a Hilbert-Schmidt
operator on the Hilbert space $H_\rho.$ The space of such sections
has a natural structure of the Hilbert space, with
$$\langle F_1,F_2\rangle = \int_{\widehat{G_\epsilon}} \tr (F_1(\rho)
F_2^*(\rho)) dP_\epsilon(\rho),$$ where $dP_\epsilon$ is the
Plancherel measure on $\widehat{G}.$ It is known that the mapping
$\mathfrak{F}_{G_\epsilon}$ is unitary and the inverse mapping is
given by the formula
$$(\mathfrak{F}_{G_\epsilon}^{-1} F)(X,A) = \int_{\widehat{G_\epsilon}} \tr (F(\rho)
\rho (X,A)^*) dP_\epsilon(\rho).$$

Now consider the {\bf quantization} mapping $$Q_\epsilon =
 \mathfrak{F}_{G_\epsilon}\, \mathfrak{F}_{G_0}^{-1}:
L_2(\widehat{G_0}, dP_0)\rightarrow L_2(\widehat{G_\epsilon},
\Lambda(G_\epsilon); dP_\epsilon).$$

This mapping is obviously unitary, it can be explicitly inverted.
The inverse mapping is called the {\bf de-quantization} or the
{\bf symbol} mapping. These mappings are closely related to Weyl
symbols, etc, see, e.g., \cite{Ho1, Ho2, BS}.

Let us note an important property of the quantization mapping: let
$1_0$ denote the function on $\widehat{G_0}$ whose value at each
point of $\widehat{G_0}$ is $1$, let $1_\epsilon$  denote the
section, whose value at each point  $\rho\in \widehat{G_\epsilon}$
is the identity operator in $H_\rho.$ One can easily see that
$$Q_\epsilon(1_0) = 1_\epsilon.$$ This implies (via the unitarity
of $Q_\epsilon$) that
$$\forall\, f \in L_1(\widehat{G_0}, dP_0)\quad
 \int_{\widehat{G_0}} f dP_0 = \langle f,1_0\rangle = \langle Q_\epsilon f,
 1_\epsilon\rangle=
\int_{\widehat{G_\epsilon}} \tr (Q_\epsilon f) dP_\epsilon.$$

We treat  functions $f:\R^4 \rightarrow \C$ as classical scalar
fields on $\R^4.$ They can also be viewed as functions on
$\widehat{G_0},$ though not integrable with respect to $dP_0.$ Let
us fix a weight function $W: SS(\R^4)\rightarrow \R_+,$ that is, a
positive function, such that $\int_{SS(\R^4)} W(\Phi) d^6 \Phi =
1.$

 The introduction of the weight
function allows one to integrate classical scalar fields with
respect to the weighted measure, and so that
$$\int_{\widehat{G_0}} f(\phi) W(\Phi) dP_0(\phi,\Phi) = \int_{\R^4} f(\phi) d^4\phi.$$

 Note that since $W$ does not depend upon
$\phi \in \R^{*4},$ then $(\mathfrak{F}_{G_0}^{-1}W)$ is supported
by the subspace $\R^4\wedge \R^4$ in $G_0 = G_\epsilon,$ and
therefore
$$(Q_\epsilon W)(\rho) = \int_{G_\epsilon} \rho(X,A)
(\mathfrak{F}_{G_0}^{-1}W)(X,A) dH(X,A)$$ $$ = \int_{\R^4\wedge
\R^4} \rho (0,A) (\mathfrak{F}_{G_0}^{-1}W)(0,A) d^6A,$$ so since
the elements $(0,A)$ belong to the center of the group
$G_\epsilon$ then the operators $\rho(0,A)$ are scalar operators,
$\rho(0,A) = \lambda (A)\id,$ so for any $\rho \in
\widehat{G_\epsilon}$ the operator $(Q_\epsilon W)(\rho)$ is
scalar. Because $W$ is real valued we also have
$$\forall\, \rho \in \widehat{G_\epsilon}\quad (Q_\epsilon W)(\rho) = (Q_\epsilon \overline{W})(\rho) = (Q_\epsilon
W)(\rho)^*.$$ So, this is a real valued scalar operator. To sum
up, $Q_\epsilon W$ can be treated as a real valued function on
$\widehat{G_\epsilon}.$ So $(Q_\epsilon W)(\rho)  = (\text {
function of } \rho) \id_\rho.$ Slightly abusing notation, we
identify this function with $Q_\epsilon W.$

Obviously,
$$\int_{\R^4} f(\phi) d^4\phi = \int_{\widehat{G_0}} f(\phi) W(\Phi) dP_0(\phi,\Phi) =
\int_{\widehat{P_\epsilon}} \tr ((Q_\epsilon f)(\rho) (Q_\epsilon
W)(\rho)) dP_\epsilon (\rho)$$ $$ = \int_{\widehat{P_\epsilon}}
\tr ((Q_\epsilon f)(\rho)) (Q_\epsilon W)(\rho) dP_\epsilon
(\rho).$$

\section{Symbol of a function of a section}

The quantization mapping establishes a one-to-one linear
correspondence between functions on $\widehat{G_0}$ and sections
of the dual bundle $\Lambda (G).$ But taking functions of
functions and functions of sections is quite different.

Let $f$ be a real-valued function on $\widehat{G_0}$. Choose $k:\R
\rightarrow \C$ be a function of one variable. Then one can
consider the function $k(f)$ as usual -- $k(f)(\phi,\Phi) =
k(f(\phi,\Phi))$, and get a new function on the same set. We treat
this obvious way of computing a function of a function as the
"commutative way".

Let $F = Q_\epsilon f$ be a section with, say, self-adjoint
values. Choose any reasonable function $k: \R \rightarrow \C,$ and
then one can define a new section $k(F)$ as follows:
$$(k(F))(\rho) = k(F(\rho)),$$ assuming that the function $k$ of
the operator $F(\rho)$ (acting on the Hilbert space $H_\rho$)
makes sense for every $\rho \in \widehat{G_\epsilon}.$

One can easily see that $$Q_\epsilon k(f) \ne k(Q_\epsilon f).$$
So $Q_\epsilon ^{-1}k(Q_\epsilon f)$ can be viewed as a different
way of computing a function of a function -- a "noncommutative
way".

As a matter of fact what we really need to compute is an action
functional which in this simplest case is defined to be
$$S_W(F) = \int_{\widehat{G_\epsilon}} \tr (k(F)(\rho)) (Q_\epsilon W)(\rho) dP_\epsilon(\rho).$$ So we try
first to compute the symbol $Q_\epsilon^{-1} k(F)$ and then use
the above mentioned fact that
$$\int_{\widehat{G_\epsilon}} \tr (k(F)(\rho))(Q_\epsilon W)(\rho) dP_\epsilon(\rho) =
\int_{\widehat{G_0}} (Q_\epsilon^{-1} k(F))(\phi,\Phi) W(\Phi)
dP_0(\phi,\Phi).$$

Since the main goal of our computations is to get a hint for our
definitions to be presented below, we are rather formal in our
computations -- we freely interchange limits and integrals, the
order of integration,  do not pay much attention to the questions
of convergence, etc. We feel that all computations below can be
made precise under some additional rather mild assumptions.

Let us first note that it is enough to compute $Q^{-1}(\exp(itF))$
for every $t\in \R$ since then one can calculate $$Q^{-1}(k(F)) =
\int_\R Q^{-1}(\exp(itF)) \hat{k}(t) dt,$$ where $\hat{k}$ is the
usual inverse Fourier transform of the function $k:$
$$k(s) = \int_\R e^{its} \hat{k}(t) dt.$$ Of course, we assume
that all integrals  make sense (this is actually a part of the
definition of a reasonable function $k$).

Let us recall several simple facts related to general Fourier
transforms:

(i) the inverse Fourier transforms of the functions $1_\epsilon,\,
\epsilon \ge 0,$ equals $\delta_0(X,Y)$ (the delta function
supported at the common neutral element of all groups in
question),

(ii) the $G_\epsilon$-convolution of functions $h_1, h_2, \cdots,
h_N$ on $G_\epsilon$ is given by the formula
$$(h_1*_\epsilon \cdots *_\epsilon h_N)(X,A) = \int_{G_\epsilon^N}
\prod_{j=1}^N h_j(X_j,A_j) \delta_{(X,A)}(\diamondsuit_{j=1}^N
(X_j,A_j)) \prod_{j=1}^N dH(X_j,A_j),$$

(iii) the Fourier transform of a $G_\epsilon$-convolution of
several functions on a group $G_\epsilon$ equals the product of
the Fourier transforms of these functions.

Using these facts plus the fact that all groups $G_\epsilon,\,
\epsilon \ge 0,$ coincide as sets, and have the same invariant
measures $dH(X,A),$ we compute

$$Q_\epsilon^{-1}(\exp(itF))(\phi,\Phi) = (\mathfrak{F}_{G_0}
\mathfrak{F}_{G_\epsilon}^{-1}(\lim_{N\rightarrow \infty}
(1_\epsilon + \frac{it F}{N})^N))(\phi,\Phi)$$
$$= \lim_{N\rightarrow \infty}(\mathfrak{F}_{G_0}((\delta_0(X,A) +
\frac{it(\mathfrak{F}_{G_\epsilon}^{-1} F)(X,A)}{N})^{*_\epsilon
N}))(\phi,\Phi)$$
$$= \lim_{N\rightarrow
\infty} \int_{G_0} \exp (i(\phi(X) + \Phi(A))) dH(X,A)\times $$
$$ \times \left(\int_{G_\epsilon^N} \prod_{j=1}^N \left(\delta_0(X_j,A_j) +
\frac{it(\mathfrak{F}_{G_\epsilon}^{-1} F)(X_j,A_j)}{N}\right)
\delta_{(X,A)}(\diamondsuit_{j=1}^N (X_j,A_j)) \prod_{j=1}^N
dH(X_j,A_j)\right)$$
$$= \lim_{N\rightarrow
\infty}\int_{G_\epsilon^N} \prod_{j=1}^N \left(\delta_0(X_j,A_j) +
\frac{it(\mathfrak{F}_{G_\epsilon}^{-1}
F)(X_j,A_j)}{N}\right)\prod_{j=1}^N dH(X_j,A_j)\times $$
$$\times \left(\int_{G_0} \exp (i(\phi(X) + \Phi(A)))
\delta_{(X,A)}(\diamondsuit_{j=1}^N (X_j,A_j)) dH(X,A)\right)$$
$$= \lim_{N\rightarrow
\infty}\int_{G_0^N} \prod_{j=1}^N \left(\delta_0(X_j,A_j) +
\frac{it(\mathfrak{F}_{G_\epsilon}^{-1}
F)(X_j,A_j)}{N}\right)\times$$ $$\times \exp
(i((\phi,\Phi)(\diamondsuit_{j=1}^N (X_j,A_j))) \prod_{j=1}^N
dH(X_j,A_j)$$
$$= \lim_{N\rightarrow
\infty}\int_{G_0^N} \prod_{j=1}^N \left(\delta_0(X_j,A_j) +
\frac{it(\mathfrak{F}_{G_\epsilon}^{-1}
F)(X_j,A_j)}{N}\right)\times $$
$$\times \, \, \overline{\exp (- i((\phi,\Phi)(\diamondsuit_{j=1}^N
(X_j,A_j)))} \prod_{j=1}^N dH(X_j,A_j)$$

Applying the Plancherel identity ( = unitarity of the Fourier
transform) for the group $G_0^N$, we see that
$$Q_\epsilon^{-1}(\exp(itF))(\phi,\Phi) = \lim_{N\rightarrow
\infty} \int_{\widehat{G_0^N}} \prod_{j=1}^N \left(1 +
\frac{it(Q_\epsilon^{-1} F)(\phi_j,\Phi_j)}{N}\right)
\prod_{j=1}^N dP_0(\phi_j,\Phi_j)\times $$ $$\times \, \overline {
\int _{G_0^N} \exp \left(i\sum_{j=1}^N \phi_j(X_j) +
\Phi_j(A_j)\right)\exp \, \left(-
i(\phi,\Phi)(\diamondsuit_{j=1}^N (X_j,A_j)\right) \prod_{j=1}^N
dH(X_j,A_j)}$$
$$ = \lim_{N\rightarrow
\infty} \int_{\widehat{G_0^N}} \prod_{j=1}^N \left(1 +
\frac{it(Q_\epsilon^{-1} F)(\phi_j,\Phi_j)}{N}\right)
\prod_{j=1}^N dP_0(\phi_j,\Phi_j)\times $$ $$\times \,  \int
_{G_0^N} \exp \left(-i\sum_{j=1}^N \phi_j(X_j) +
\Phi_j(A_j)\right)\exp \, \left(i(\phi,\Phi)(\diamondsuit_{j=1}^N
(X_j,A_j)\right) \prod_{j=1}^N dH(X_j,A_j).$$

\subsection{Simplifications}
We need to simplify the term in the last line:
$$\widetilde{\Psi_N}(\phi; \Phi;
\phi_1,\cdots,\phi_N;\Phi_1,\cdots, \Phi_N)$$ $$ = \int _{G_0^N}
\exp \left(-i\sum_{j=1}^N \phi_j(X_j) + \Phi_j(A_j)\right)\exp \,
\left(i(\phi,\Phi)(\diamondsuit_{j=1}^N (X_j,A_j)\right)
\prod_{j=1}^N dH(X_j,A_j).$$

Let us perform the following substitution in this integral:
$$\diamondsuit_{j=1}^k (X_j,A_j) = (Y_k,B_k),\, \, k =
1,2,\cdots,N,\quad \text { let } (Y_0,B_0) = (0,0)$$ Then
$$(Y_{k-1},B_{k-1}) \diamondsuit (X_k,A_k) = (Y_k,B_k),$$ so
$$(X_k,A_k) = (Y_k,B_k)\diamondsuit (-Y_{k-1},-B_{k-1}) = (Y_k -
Y_{k-1}, A_k - A_{k-1} - \frac {\epsilon}{2} Y_k\wedge Y_{k-1}),$$
$$ k = 1,2,\cdots, N.$$ So the substitution is one-to one, its
Jacobian is apparently $1$, so we get
$$\widetilde{\Psi_N}(\phi; \Phi;
\phi_1,\cdots,\phi_N;\Phi_1,\cdots, \Phi_N)$$
$$= \int _{G_0^N}\exp \, (-i)\left(\sum_{j=1}^N \phi_j(Y_j - Y_{j-1}) + \Phi_j(B_j - B_{j-1} - \frac
{\epsilon}{2} Y_j\wedge Y_{j-1})\right)\times $$ $$\times \exp \,
i(\phi (Y_N) + \Phi (B_N)) \, \, \prod_{j=1}^N dH(Y_j,B_j)$$
$$= \int_{\R^{*6N}} \exp \, i\left(   (\Phi - \Phi_N)(B_N) +
\sum_{j=1}^{N-1} (\Phi_{j+1} -
\Phi_j)(B_j))\right)\frac{\prod_{j=1}^N d^6B_j}{(2\pi)^{6N}}\times
$$ $$\times \int_{\R^{*4N}} \exp \, i \left((\phi - \phi_N)(Y_N) +
\sum_{j=1}^{N-1} (\phi_{j+1} - \phi_j)(Y_j) + \frac {\epsilon}{2}
\sum_{j=1}^N \Phi_j(Y_j\wedge Y_{j-1})\right)\frac{\prod_{j=1}^N
d^4Y_j}{(2\pi)^{4N}}
$$
$$= \delta_0(\Phi - \Phi_N) \prod_{j=1}^{N-1} \delta_0(\Phi_{j+1} -
\Phi_j)\times $$ $$\times \int_{\R^{*4N}} \exp \, i\left( (\phi -
\phi_N)(Y_N) + \sum_{j=1}^{N-1} (\phi_{j+1} - \phi_j)(Y_j) + \frac
{\epsilon}{2} \Phi\left(\sum_{j=1}^N Y_j\wedge
Y_{j-1}\right)\right)\frac{\prod_{j=1}^N d^4Y_j}{(2\pi)^{4N}}.
$$

So, letting $\phi_{N+1} = \phi, \Phi_{N+1} = \Phi,$ if necessary,
we see that
$$\widetilde{\Psi_N}(\phi; \Phi;
\phi_1,\cdots,\phi_N;\Phi_1,\cdots, \Phi_N)$$ $$=
 \prod_{j=1}^{N} \delta_0(\Phi - \Phi_j) \int_{\R^{*4N}}
\exp \, i\left(\sum_{j=1}^{N} (\phi_{j+1} - \phi_j)(Y_j)\right)
\exp \, i \Phi\left(\frac {\epsilon}{2}\sum_{j=1}^N Y_j\wedge
Y_{j-1}\right)\frac{\prod_{j=1}^N d^4Y_j}{(2\pi)^{4N}}
$$

\section{Computation of the simplest action functional} Now let
us assume that the function $f$ does not depend upon $\Phi,$ i.e.,
$f$ is a classical scalar field. Let us now compute the action
functional
$$S_W(Q_\epsilon^{-1}(\exp itQ_\epsilon
f)) = \int_{\widehat{G_0}} Q_\epsilon^{-1}(\exp itQ_\epsilon
f)(\phi,\Phi) W(\Phi) dP_0(\phi,\Phi)$$
$$ = \lim_{N\rightarrow \infty}\int_{\widehat{G_0}}W(\Phi)
dP_0(\phi,\Phi) \int_{\R^{*4N}} \prod_{j=1}^N \left(1 + \frac{it
f(\phi_j)}{N}\right) \frac{\prod_{j=1}^N
d^4\phi_j}{(2\pi)^{4N}}\times $$
$$\times \int_{\R^{4N}}
\exp \, i\left(\sum_{j=1}^{N} (\phi_{j+1} - \phi_j)(Y_j)\right)
\exp \, i \Phi\left(\frac {\epsilon}{2}\sum_{j=1}^N Y_j\wedge
Y_{j-1}\right)\frac{\prod_{j=1}^N d^4Y_j}{(2\pi)^{4N}}
$$
$$ = \lim_{N\rightarrow \infty}\int_{\R^{*4N}} \prod_{j=1}^N \left(1 +
\frac{it f(\phi_j)}{N}\right) \frac{\prod_{j=1}^N
d^4\phi_j}{(2\pi)^{4N}}\times $$
$$\times \int_{\R^{*4}} \frac{d^4\phi_{N+1}}{(2\pi)^4} \int_{\R^{4N}}
\exp \, i\left(\sum_{j=1}^{N} (\phi_{j+1} -
\phi_j)(Y_j)\right)\frac{\prod_{j=1}^N d^4Y_j}{(2\pi)^{4N}}\times
$$
$$\times \int_{SS(\R^4)} W(\Phi)\exp \, i \Phi\left(\frac {\epsilon}{2}\sum_{j=1}^N Y_j\wedge
Y_{j-1}\right)\frac{d^6\Phi}{(2\pi)^6}.$$

For $A \in \R^4\wedge \R^4$ denote $$w(A) = \int_{SS(\R^4)}
W(\Phi)\exp \, i \Phi (A)\frac{d^6\Phi}{(2\pi)^6}.$$  Obviously,
$$w(0) = \int_{SS(\R^4)} W(\Phi)\frac{d^6\Phi}{(2\pi)^6} = 1.$$

Then we get
$$S_W(Q_\epsilon^{-1}(\exp itQ_\epsilon
f)) = \lim_{N\rightarrow \infty}\int_{\R^{*4N}} \prod_{j=1}^N
\left(1 + \frac{it f(\phi_j)}{N}\right) \frac{\prod_{j=1}^N
d^4\phi_j}{(2\pi)^{4N}}\times $$
$$\times \int_{\R^{*4}} \frac{d^4\phi_{N+1}}{(2\pi)^4} \int_{\R^{4N}}
\exp \, i\left(\sum_{j=1}^{N} (\phi_{j+1} - \phi_j)(Y_j)\right)
w\left(\frac {\epsilon}{2}\sum_{j=1}^N Y_j\wedge Y_{j-1}\right)
\frac{\prod_{j=1}^N d^4Y_j}{(2\pi)^{4N}}.
$$
Integrating with respect to $\phi_{N+1}$ we get a factor of
$\delta_0(Y_N)$ and therefore we finally obtain
$$S_W(Q_\epsilon^{-1}(\exp itQ_\epsilon
f)) = \lim_{N\rightarrow \infty}\int_{\R^{*4N}} \prod_{j=1}^N
\left(1 + \frac{it f(\phi_j)}{N}\right) \frac{\prod_{j=1}^N
d^4\phi_j}{(2\pi)^{4N}}\times $$
$$\times \int_{\R^{4(N-1)}}
\exp \, i\left(\sum_{j=1}^{N-1} (\phi_{j+1} - \phi_j)(Y_j)\right)
w\left(\frac {\epsilon}{2}\sum_{j=1}^{N-1} Y_j\wedge
Y_{j-1}\right) \frac{\prod_{j=1}^{N-1} d^4Y_j}{(2\pi)^{4(N-1)}}.
$$

If $\epsilon \rightarrow 0,$ then the function $w\left(\frac
{\epsilon}{2}\sum_{j=1}^{N-1} Y_j\wedge Y_{j-1}\right)$ goes to
the constant function equal to $w(0) = 1,$ the Fourier transform
of this function goes to $\delta_0,$ so
$$\int_{\R^{4(N-1)}}
\exp \, i\left(\sum_{j=1}^{N-1} (\phi_{j+1} - \phi_j)(Y_j)\right)
w\left(\frac {\epsilon}{2}\sum_{j=1}^{N-1} Y_j\wedge
Y_{j-1}\right) \frac{\prod_{j=1}^{N-1}
d^4Y_j}{(2\pi)^{4(N-1)}}\rightarrow \prod_{j=1}^N \delta
(\phi_{j+1} - \phi_j),$$ and so $$\lim_{\epsilon \rightarrow
0}S_W(Q_\epsilon^{-1}(\exp itQ_\epsilon f)) = \lim_{N\rightarrow
\infty}\int_{\R^{*4N}} \prod_{j=1}^N \left(1 + \frac{it
f(\phi_j)}{N}\right)\prod_{j=1}^N \delta (\phi_{j+1} - \phi_j)
\frac{\prod_{j=1}^N d^4\phi_j}{(2\pi)^{4N}}$$
$$=\lim_{N\rightarrow
\infty}\int_{\R^{*4}} \left(1 + \frac{it f(\phi)}{N}\right)^N
\frac{d^4\phi}{(2\pi)^4}= \int_{\R^{*4}} \exp (itf(\phi))
\frac{d^4\phi}{(2\pi)^4},
$$ which is what one should have expected.

\subsection{Formal passage to the limit as $N\rightarrow \infty$}
Now let us extract some hints from this formula. Since
these are going to be only hints (or, better to say, motivations
for our subsequent definitions) we proceed very formally.

Introduce a mapping $\phi: [0,1]\rightarrow \R^{*4},$ and let
$\phi_j = \phi(j/N),\, \, j = 1,2,\cdots,N.$ The set of such
mappings is denoted $T(\R^{*4}).$ Also, introduce a mapping
$Z:[0,1]\rightarrow \R^4,$ and let $Z_j = Z(j/N),\, j =
1,2,\cdots, N-1.$ The set of such mappings is denoted $T(\R^4).$

Then
$$S_W(Q_\epsilon^{-1}(\exp itQ_\epsilon f)) = \lim_{N\rightarrow
\infty}\int_{\R^{*4N}} \exp it\left( \sum_{j=1}^N \frac1N
f(\phi(j/N))\right) \frac{\prod_{j=1}^N
d^4\phi(j/N)}{(2\pi)^{4N}}\times $$
$$\times \int_{\R^{4(N-1)}}
\exp \, i\left(\sum_{j=1}^{N-1} \frac{\phi((j+1)/N) -
\phi(j/N)}{1/N}\,  Z(j/N) \frac 1N\right)\times $$ $$\times
w\left(\frac {\epsilon}{2}\sum_{j=1}^{N-1}
\frac{Z(j/N)-Z((j-1)/N)}{1/N}\wedge Z((j-1)/N)\frac 1N\right)
\frac{\prod_{j=1}^{N-1} d^4Z(j/N)}{(2\pi)^{4(N-1)}}.
$$
$$=\int_{T(\R^{*4})} \exp \left( it \int_0^1 f(\phi(\sigma))d\sigma\right)
D\phi(\cdot)\times $$ $$\times \int_{T(\R^4)} \exp \left(i\int_0^1
\phi'(\sigma)(Z(\sigma))d\sigma\right) w\left(\int_0^1
\frac{\epsilon}2 Z'(\sigma)\wedge Z(\sigma)d\sigma\right)
DZ(\cdot).$$

\subsection{From parametrized paths to measures}
A continuous parametrized path $\phi: [0,1] \rightarrow \R^{*4}$
can have an almost arbitrary form (in particular, it can fill a
whole cube in $\R^{*4}$, due to the famous Peano example), we
prefer to treat a parametrized curve $\phi$ as a probability
measure on $\R^{*4},$ supported on (the closure of) the range of
$\phi,$ with the measure of a piece of the curve equal to the one
dimensional measure of its pre-image.
 Then $\int_0^1 f(\phi(\sigma))d\sigma$ is simply the
integral of $f$ against this measure, we denote it $\langle
f,\phi\rangle.$

Accordingly, the space $T(\R^{*4}) $ can be treated  as the space
of probability measures on $\R^{*4}.$ Let us drop the condition
that $\langle 1,\phi\rangle = 1$, instead we require that the
measure $\phi\in T(\R^{*4})$ is nonnegative and $\langle
1,\phi\rangle \le 1.$

We would like to treat expression $$D\phi(\cdot) \int_{T(\R^4)}
\exp \left(i\int_0^1 \phi'(\sigma)(Z(\sigma))d\sigma\right)
w\left(\int_0^1 \frac{\epsilon}2 Z'(\sigma)\wedge
Z(\sigma)d\sigma\right) DZ(\cdot)$$  as a measure
 on the (weakly compact) set
$T(\R^{*4})$ of  measures on $\R^{*4}.$ However, this is not
likely, e.g., this expression is explicitly complex-valued, which
causes a lot of difficulties. However, the infinite-dimensional
integral is over the space of parametrized curves in $\R^4$, but
many of these curves lead to the same measures on $T(\R^{4}).$ On
the other hand the expression $k(\langle f,\phi\rangle$ depends
upon a measure. We presume (though have not yet proven this) that
a "symmetrized" version of the expression $D\phi(\cdot)
\int_{T(\R^4)} \exp \left(i\int_0^1
\phi'(\sigma)(Z(\sigma))d\sigma\right) w\left(\int_0^1
\frac{\epsilon}2 Z'(\sigma)\wedge Z(\sigma)d\sigma\right)
DZ(\cdot)$ obtained by summation over all parametrized curves
leading to the same measure, is likely to define a positive
measure on the space $T(\R^{*4}).$ We have some promising
developments in the finite-dimensional case associated with finite
analogues of the groups in question.

Let $dM_{W,\epsilon}(\phi)$ denote the hypothetical positive
measure on the space $T(\R^{*4})$ such that
$$S_W(Q_\epsilon^{-1}(\exp itQ_\epsilon f)) = \int_{T(\R^{*4})}
\exp (it\langle f,\phi\rangle) dM_{W,\epsilon}(\phi).$$

Keeping in mind that any function $k: \R \rightarrow \C$ is an
integral over exponentials, we see that for any function $k$ we
have $$S_W(Q_\epsilon^{-1}(k(Q_\epsilon f))) = \int_{T(\R^{*4})} k
\left( \langle f,\phi\rangle\right) dM_{W,\epsilon}(\phi).$$

One can easily show that a natural extension of this formula holds
for vector valued fields $\vec{f}: \R^4 \rightarrow \R^m,$ and for
functions $k: \R^k \rightarrow \C:$
$$S_W(Q_\epsilon^{-1}(k(Q_\epsilon \vec{f}))) =
\int_{\widehat{G_\epsilon}} \tr (k(Q_\epsilon \vec{f}))Q_\epsilon W dP_\epsilon =
 \int _{T(\R^{*4})} k(\langle \vec{f}, \phi \rangle )dM_{W,\epsilon}(\phi),$$
 where $ Q_\epsilon \vec{f} = (Q_\epsilon f_i)_{i=1}^m,$ the function $k(F_1,\cdots,F_m)$ of several operators is
 defined in the symmetric (Weyl) way (one first defines the exponential functions
 $k = \exp i\sum_{i=1}^m t_i x_i$ of the operators -- the problem of ordering does not exist for these functions,
 then any function is represented as an integral over exponentials),
 $\langle \vec{f}, \phi \rangle \in \R^{*4}.$

\section{Noncommutative space-time and fields on it} A vector
field  $\vec{f}$ on $\R^{*4}$ gives rise to a {\bf linear} vector
valued function $\phi \rightarrow \langle \vec{f},\phi \rangle$ on
$T(\R^{*4}).$ Then we take a scalar function of this linear vector
valued function and integrate it over the space-time to obtain the
quantities needed to form the action functional.

So if we take the classical vector fields on $\R^{*4}$, transplant
them to $\widehat{G_\epsilon},$ so that when computing a function
$k$ of the field we are able to take the noncommutativity of the
space-time into account, and then form the noncommutative version
of the action functional of such classical fields, we arrive at an
infinite-dimensional integral of a scalar function of a {\bf
linear } vector valued function on $T(\R^{*4}).$ However, there is
a serious problem arising here: let the function $k$ be invariant
under action of a subgroup $\mathfrak{G} \subset GL(m):$
 $$k(x)= k(gx),\, \, \forall\, g\in \mathfrak{G},\, x\in \R^m.$$ Then the function $k(\vec{f})$
 is invariant under the gauge group of continuous mappings from $\R^{*4}$ to $\mathfrak{G}$, i.e.,
$k(\vec{f}(\psi)) =  k(g(\psi)\vec{f}(\psi)),\, \, \forall\, \psi
\in \R^{*4}\, \,  g(\psi) \in \mathfrak{G}.$ Apparently the
function $k(\langle \vec{f},\phi\rangle)$ on $T(\R^{*4})$ is not
invariant under this gauge group. However, this function
$k(\langle \vec{f},\phi\rangle)$ is obviously invariant under the
action of the {\bf extended gauge group} consisting of continuous
mappings from $T(\R^{*4})$ to $\mathfrak{G}$ (the set $T(\R^{*4})$
is endowed with the weak topology):
$$k(\langle \vec{f},\phi\rangle) = k(g(\phi)\langle
\vec{f},\phi\rangle),\, \, \forall\, \phi \in T(\R^{*4})\, \, \,
g(\phi)\in \mathfrak{G}.$$ But in this case the vector valued
function $g(\mu)\langle \vec{f},\phi\rangle$ is {\bf not linear}
in $\phi.$

Restricting ourselves only to linear vector valued functions $\phi
\mapsto  \langle \vec{f},\phi\rangle$ was caused by the fact that
we were considering only usual vector fields $\vec{f}$ on the
space-time $\R^{*4}.$ The extended gauge transformations produce
nonlinear vector valued functions on $T(\R^{*4}),$ which obviously
do not come from the classical vector fields on $\R^{*4}.$

In quantum field theory the quantities of interest arise when one
 integrates the exponentials of the action functional (as a function of a
field) over all fields. So a field is a "silent" variable, being
integrated over, and therefore it seems to be not a big deal if we
broaden the set of fields.

Let us very significantly broaden the notion of a vector field in
this (noncommutative) context, allowing any (not necessarily
linear) vector valued continuous function on the weakly compact
set of  measures $T(\R^{*4}).$ The extended gauge group naturally
acts on such vector fields. The action functional associated with
such field is defined as
$$S_W(k(\vec{f})) = \int _{T(\R^{*4})} k(\vec{f}(\phi))dM_{W,\epsilon}(\phi).$$

So we define

{\bf the noncommutative space-time = the set $T(\R^{*4})$ of
measures on the commutative space-time $\R^{*4}.$}

the noncommutative space-time $T(\R^{*4})$ is a weakly compact
subset in an infinite-dimensional dual Banach space. The usual
space-time $\R^{*4}$ is imbedded into the noncommutative
space-time by the mapping $\R^{*4} \ni \psi \mapsto \delta_\psi =
\phi \in T(\R^{*4}).$ Since the delta measures form  kind of a
basis in the linear space of measures on $\R^{*4},$ then a
classical scalar field is an arbitrary function on this basis, and
it naturally extends to the whole linear space of measures as a
linear function. However, there exist many  extensions of the same
function on the basis to nonlinear functions on the whole space of
measures. These nonlinear functions are not distinguishable from
linear functions if we observe only their values on the basis.
Since for a very small $\epsilon$ our space-time is almost
commutative, which means that the support of the measure
$dM_{W,\epsilon}(\phi)$  is very close to the set of delta
measures, then all nonlinear extensions are indistinguishable from
linear extensions, and we in fact are able to observe only
functions on the commutative space-time. In other words, the
classical fields on the commutative space-time $\R^{*4}$ are only
shadows of fields on the noncommutative space-time $T(\R^{*4})$,
and for sufficiently small $\epsilon$ we observe only these
shadows.

\section{Geometry on the noncommutative space-time}

\subsection{Tangent and cotangent bundles over the noncommutative
space-time} By our definition, the noncommutative space-time
$T(\R^{*4})$ is a simplex in the infinite dimensional dual Banach
space $\mathfrak{M}_4$ of  measures (not necessarily positive, and
of any total variation) on $\R^{*4}$. This space $\mathfrak{M}_4$
is dual to the space $C_0(\R^{*4})$ of continuous compactly
supported functions on $\R^{*4}.$

As usual, the tangent space to a linear space at any point
coincides with this linear space. So, the tangent space
$\mathfrak{T}_\phi T(\R^{*4})$ to $T(\R^{*4})$ at each point $\phi
\in T(\R^{*4})$ is simply $\mathfrak{M}_4.$ So the tangent bundle
$\mathfrak{T}T(\R^{*4})$ is simply $T(\R^{*4})\times
\mathfrak{M}_4.$

Therefore a tangent vector field $\Delta$ on $T(\R^{*4})$ -- a
section of the tangent bundle -- is a rule assigning a measure
$\nu \in \mathfrak{M}_4$ to each $\phi \in T(\R^{*4}):\, \, \,
\Delta(\phi) = \nu.$

Let $A_0$ denote the algebra of complex-valued functions on
$T(\R^{*4}).$

If $f\in A_0$  then we define its derivative at a point $\phi\in
T(\R^{*4})$ with respect to a tangent vector $\nu \in
\mathfrak{T}_\phi T(\R^{*4})$ in the usual way:
$$(\partial_\nu f)(\phi) = \lim_{t\rightarrow 0} \frac 1t (f(\phi
+ t\nu) - f(\phi)).$$

We say that $f$ is weakly Frechet differentiable at $\phi$ if
there exists a function $df_\phi \in C_0(\R^{*4})$ such that for
any $\nu \in \mathfrak{T}_\phi T(\R^{*4})$ we have
$$(\partial_\nu f)(\phi) = \langle df_\phi,\nu\rangle.$$

Therefore we define
 $\mathfrak{T}^*_\phi T(\R^{*4})$ -- the cotangent
space to $T(\R^{*4})$ at $\phi \in T(\R^{*4})$ -- as
$C_0(\R^{*4})$ (it is a pre-dual to the tangent space, rather than
the dual).

A $1$-form (or a cotangent vector field) on $T(\R^{*4})$ is a
section of the cotangent bundle, i.e, a rule assigning a function
$h_\phi \in C_0(\R^{*4})$ to each $\phi \in T(\R^{*4}).$ Pairing a
cotangent vector field $h$ and a tangent vector field $\nu$ yields
a function on $T(\R^{*4}):$
$$\langle h,\nu\rangle (\phi) = \langle h_\phi,\nu_\phi\rangle.$$

For any tangent vector field $\Delta$ on $T(\R^{*4})$ we define an
operator on $A_0$, which we continue to denote $\Delta$ despite an
obvious abuse of notation:
$$(\Delta f)(\phi) = (\partial_{\Delta(\phi)} f)(\phi).$$
For a weakly Frechet differentiable function $f$ we have
$$(\Delta f)(\phi) = \langle df_\phi,\Delta(\phi)\rangle.$$
This operator is obviously linear, it also satisfies the Leibniz
rule, so it is a {\bf local differentiation} of the algebra $A_0$
(locality means that the support of $\Delta(f)$ is contained in
the support of $f$). Therefore the space of tangent vector fields
is also denoted $Diff A_0$.

\subsection{Gauge fields on the noncommutative space-time}
There is a natural basis in the linear space $\mathfrak{M}_4$ --
it consists of all $\delta$ measures, $\delta_x,\, x\in \R^{*4}.$
For each $x\in \R^{*4}$ we let $\partial_x$ denote differentiation
with respect to $\delta_x,$ it can be viewed as a (translation
invariant) tangent vector field on $T(\R^{*4}).$ One can show that
$$\forall\, x,y \in \R^{*4},\, x\ne y, \, \, [\partial_x,
\partial_y] = 0.$$

Let $E = T(\R^{*4})\times \C^m$ be the trivial vector bundle over
the noncommutative space-time. Let $$\Gamma(E)= \{ \vec{f} =
(f_i)_{i=1}^m: \forall\, i, 1\le i\le m\, \,
f_i:T(\R^{*4})\rightarrow \C\}$$ denote the space of its sections
-- mappings from $T(\R^{*4})$ to $\C^m.$ Then  a tangent vector
field $\Delta$ still defines an operator on $\Gamma(E):$
$$\Delta \vec{f} = (\Delta f_i)_{i=1}^m.$$
In particular, standard tangent vector fields $\partial_x$ define
operators on $\Gamma(E).$ For each $x\in \R^{*4}$ and for each
$\phi \in T(\R^{*4})$  choose a linear operator $A_x(\phi): \C^m
\rightarrow \C^m.$ Consider the operators $$\nabla_x = \partial_x
+ A_x(\phi)$$ on $\Gamma(E):$ $$(\nabla_x \vec{f})(\phi) =
(\partial_x \vec{f})(\phi) + A_x(\phi)\vec{f}(\phi).$$ They are
called the {\bf covariant differentiations, or connections, or
gauge fields}.

Easy to see that
$$[\nabla_x,\nabla_y] = \partial_x A_y - \partial_y A_x +
[A_x,A_y],$$ i.e., $([\nabla_x,\nabla_y]\vec{f})(\phi)$ is
multiplication by an operator in $\C^m.$ This operator is denoted
$F_{xy}(\phi)$ and is called the {\bf curvature of the connection}
$\nabla$.

Let $\mathfrak{G}$ be a Lie group and let $\rho$ be its
irreducible representation on $\C^m.$ Let $\gamma$ be its Lie
algebra. We may consider the connections such that $\forall\, x\in
\R^{*4}, \forall\, \phi \in T(\R^{*4}) \, \, A_x(\phi) \in
\rho'(\gamma).$ Such covariant differentiations help ensure that
the related expressions are gauge invariant, with the gauge group
$\mathfrak{G}.$ This means that the expressions are invariant
under the transformations $\vec{f}(\phi) \mapsto
g(\phi)\vec{f}(\phi)$ for an arbitrary function $\phi \mapsto
g(\phi)\in \mathfrak{G}.$

We are mostly interested in  the following types of expressions:
$$K_1(\nabla; \phi) = \int_{\R^{*16}} \langle F_{xy}(\phi), F_{zw}(\phi)\rangle_\gamma B(x,z)
B(y,w) d^4xd^4yd^4zd^4w,$$ where
$\langle\cdot,\cdot\rangle_\gamma$ is the Killing form -- a
$\mathfrak{G}$-invariant bilinear form on $\gamma,$
$B(\cdot,\cdot)$ is a Lorentz invariant function;

$$K_2(\nabla, \vec{f}; \phi) = \int_{\R^{*8}} \langle (\nabla_x \vec{f})(\phi), (\nabla_y
\vec{f})(\phi)\rangle _m B(x,y) d^4xd^4y,$$ where $\langle
\cdot,\cdot\rangle_m$ is a $\mathfrak{G}$-invariant bilinear form
on $\C^m.$

Now we can construct the main object of the quantum gauge field
theory on the noncommutative space time:

$$W[\vec{J}] = \int Df D\nabla \times $$ $$\times \exp i\int_{T(\R^{*4})} \left(
K_1(\nabla; \phi) + K_2(\nabla, \vec{f}; \phi) + V(\vec{f}(\phi))
- \langle \vec{J}(\phi),\vec{f}(\phi)\rangle \right)
dM_{W,\epsilon}(\phi).$$

We believe that it is  possible to develop  Feynman rules for
computation of this integral, since they are quite algebraic and
do not depend too much on the finite dimensionality of the
classical. We hope to deal with this problem in future
publications.

\end{document}